\documentclass{emulateapj}
\usepackage{graphicx}
\usepackage[none]{hyphenat}
\usepackage{amsmath}
\usepackage{booktabs}
\usepackage{multirow}
\usepackage{txfonts}
\usepackage{enumitem}
\usepackage{amsmath}
\usepackage{algorithmic}
\usepackage{bm}
\usepackage{mathrsfs}
\usepackage{color}
\usepackage{natbib}
\usepackage[colorlinks,linkcolor=blue,anchorcolor=blue,citecolor=blue]{hyperref}

\begin{document}

\title{Studying the Intermittency of Magnetized Interstellar Medium by Synchrotron Polarization Intensity and its Gradient}
\author{Ru-Yue Wang\altaffilmark{1}, Jian-Fu Zhang\altaffilmark{2,3}}
\email{jfzhang@xtu.edu.cn, rywang@gxmzu.edu.cn}
\altaffiltext{1}{School of Physics and Electronic Information, Guangxi Minzu University, Nanning 530006, People's Republic of China}
\altaffiltext{2}{Department of Physics, Xiangtan University, Xiangtan, Hunan 411105, People's Republic of China}
\altaffiltext{3}{Key Laboratory of Stars and Interstellar Medium, Xiangtan University, Xiangtan, Hunan 411105, People's Republic of China}

\begin{abstract}
\centering

Based on synthetic polarization observations, we qualitatively investigate the intermittency of the magnetized interstellar medium (ISM) through synchrotron polarization intensity (SPI) and synchrotron polarization gradient (SPG). Three principal findings emerge from our analysis: (1) The intermittency of the magnetized ISM is strongly modulated by magnetization and compressibility, with pronounced modulation effects predominantly observed in regimes characterized by strong magnetization and high compressibility. (2) The characterized intermittency of the magnetized ISM demonstrates a dependence on the electron spectral index under conditions where the random magnetic field dominates. (3) SPG enables the identification of finer three-dimensional fractal structures in the case of strong Faraday depolarization. The synergistic application of SPI and SPG substantially enhances the reliability of probing magnetized ISM intermittency from real observations.
\end{abstract}


\section{Introduction} 
Magnetohydrodynamic (MHD) turbulence is ubiquitous in astrophysical environments \citep{Armstrong1995, Elmegreen2004} and exerts a regulatory influence on key astrophysical processes, including star formation \citep{MacLow2004, McKee2007}, acceleration and propagation of cosmic rays \citep{Yan2008ApJ684, Yan2008ASPC385}, angular momentum transfer \citep{Sano2004, Pessah2010}, heat conduction \citep{Narayan2001}, and turbulent magnetic reconnection \citep{Lazarian1999}. These processes are closely linked to the inherent properties of MHD turbulence, namely anisotropy, compressibility and intermittency. While the anisotropy and compressibility of MHD turbulence have been studied in great detail \citep{wang2020, Wang2022, Zhang2022FrASS}, relatively little attention has been devoted to its intermittency. Therefore, further in-depth exploration is needed.

Intermittency plays an important role in acceleration and propagation of cosmic rays. \cite{Butsky2024} proposed that cosmic rays are scattered by intermittent structures with small volume-filling factors.
However, the diffusion is unaffected by the spatial intermittency of magnetic field \citep{chen2020}. 
\cite{Zelenyi2011} also demonstrated that enhanced intermittency leads to the increase in acceleration efficiency. Furthermore, the intermittency is related to energy cascade.  \cite{Maron2001ApJ} obtained shallow spectra of $-3/2$, which has been confirmed by \cite{Yang2017ApJ}. 

The earliest landmark work on turbulence can be traced back to Kolmogorov theory \citep{Kolmogorov1941}. This theory mainly predicted two power-law relationships, one in wavevector space and the other in real space. The former corresponds to the well-known energy spectrum $E(k)\sim k^{-5/3}$, whereas the latter provides the relation between the multi-order structure function of velocity fluctuations and the spatial scale, i.e., $SF^{(p)} \propto r^{\zeta(p)}$, with the scaling exponent ${\zeta(p)}=p/3$.
It should be noted that this theory does not incorporate spatial fluctuations of local dissipation rate \citep{Frisch1995}. In reality, however, MHD turbulence is intrinsically intermittent.
Subsequent theories have provided further refinements and put forward intermittency models to reveal dissipative structures \citep{She1994, muller2000}.

The intermittency of MHD turbulence has long been extensively investigated by numerical simulations. Given that essential properties of MHD turbulence are embodied in the density, velocity, and magnetic field, researches into its intermittency have primarily focused on these three quantities. Density, as one of the most readily available statistics from observations, exhibits intermittency that depends on the driving mechanisms, particularly for compressive forcing \citep{Federrath2010, Cho2022, Beattie2022}. Furthermore, Alfv\'enic and sonic Mach numbers are key parameters influencing the intermittency of density. \cite{Kowal2007} pointed out that the intermittency becomes more pronounced in the super-Alfv\'enic or supersonic turbulence case. 
This holds for the 3D velocity, 
despite the intrinsic differences between the two fields.
The intermittency of velocity is further affected by the reference frame, and the scaling exponent depends on the direction relative to local magnetic field  \citep{Kowal2010}. 
In the reduced MHD turbulence, the dynamical coupling between velocity and magnetic field is commonly considered, and $\rm Els\ddot asser$ fields are adopted to investigate their intermittency \citep{Chandran2015}.
Compared with the velocity, magnetic field exhibits stronger intermittency \citep{Cho2003ApJ, Haugen2004}. 
\cite{Gao2026} have investigated magnetic intermittency in relativistic MHD turbulence, revealing that strong turbulence exhibits more pronounced intermittency than weak turbulence.

Synthetic observations based on numerical simulations have further advanced our understanding of MHD turbulence. In these studies, the intermittency of column density exhibits a dependence on sonic Mach number and can thus be used to constrain this quantity \citep{Kowal2007}. 
In general, column density reveals weaker intermittency than its three-dimensional counterpart. This conclusion also applies to other 2D statistics, such as centroid velocity and SPI. Nonetheless, the intermittency of these two quantities is also affected by other factors. 
\cite{Federrath2010} obtained that compressive forcing leads to stronger intermittency in centroid velocity, which is consistent with the finding for velocity.
Meanwhile, Faraday depolarization serves as a key factor influencing the intermittency of SPI \citep{Wang2024}. 
The intermittency can also inversely regulate other properties of MHD turbulence, such as anisotropy. \cite{Ho2021} proposed that the local clustering of fast modes can cause the anisotropy of velocity gradients to change direction. 

More significantly, extensive studies on intermittency have been performed in real astrophysical environments. In the solar wind, the intermittency of turbulence is related to local heating processes \citep{Osman2011}. 
As discussed by \cite{Phillips2023}, inhomogeneous heating of electrons arises from dissipation of turbulent fluctuations near intermittent structures. 
Furthermore, \cite{Wang2015} found that the region near magnetic reconnection sites exhibits stronger intermittency in the ion dissipation range compared to the ambient solar wind turbulence.
Apart from the solar wind, electron heating in the Satur$\rm n'$s magnetosphere  has been verified to be closely associated with intermittent structures \citep{Xu2023}.
For the diffuse molecular gas, \cite{Hily-Blant2008} demonstrated a more pronounced intermittency in Polaris field than in Taurus one. Based on intermittency characteristics, the low-latitude region of Galactic ISM is predicted to be in the sub-Alfv\'enic and supersonic turbulence regime \citep{Wang2024}.

Under the assumption of an isotropic electron distribution, \cite{Lazarian2012} predicted that the electron spectral index merely alters the amplitude of synchrotron fluctuations without affecting the recovered scaling slope of underlying MHD turbulence (see the footnote in \cite{Zhang2018ApJ} for numerical verification). In this work, we first explore whether variations in the electron power-law index can influence the intermittency of the magnetized ISM. We further investigate the capability of SPG statistics in revealing the intermittent properties of magnetized ISM. This paper is organized as follows. In Section \ref{sec:theo-meth}, we introduce theoretical models of intermittency and synchrotron radiative processes, as well as methods used to characterize intermittency. Section \ref{sec:simulation} describes details of numerical simulations. In Section \ref{sec:result}, we show numerical results regarding intermittency. The discussion and summary are provided in Sections \ref{sec:discussion} and \ref{sec:summary}, respectively.

\section{Theories and methods} \label{sec:theo-meth}
\subsection{Theoretical Models of Intermittency}

Early theory assumed that the cascade process of incompressible hydrodynamic turbulence is self-similar, implying that its energy spectrum follows a power-law distribution \citep{Kolmogorov1941}.
The scaling behavior is also manifested in structure functions: the multi-order structure functions of velocity fluctuations scale with separation length as described by the following relation:  
\begin{equation}
\langle |\delta v| ^{p} \rangle \propto l^{\zeta(p)},
\end{equation}
where the relation between scaling exponent and order follows $\zeta(p)=p/3$.
However, subsequent studies have confirmed that the inertial range does not exhibit strict self-similarity.
This manifests that the scaling exponents of structure function gradually depart from $p/3$ as the order $p$ increases.
Latter, the most commonly used models have been proposed to explain this phenomenon, expressed by the following formula:
\begin{equation}
\zeta(p)=\frac{p}{g}(1-x)+C[1-(1-x/C)^{p/g}],
\end{equation}
where velocity scales as $v_{l}\sim l
^{1/g}$, energy cascade rate scales as $t_{l}^{-1}\sim l^{-x}$, and $C=3-D$ represents the codimension of dissipative structure with the dimension $D$. In hydrodynamic turbulence, Kolmogorov scaling gives $g=3$ and $x=\frac{2}{3}$. The codimensions $C=2$ for 
one-dimensional dissipative structure and $C=1$ for two-dimensional counterparts correspond to the SL \citep{She1994} and MB \citep{muller2000} theoretical models, respectively.

\subsection{Synchrotron Radiative Processes}
Relativistic electron and magnetic field are two key factors responsible for generation of synchrotron radiation. Since synchrotron radiation intensity and polarization intensity can provide key information on magnetic field, these two statistics are used to explore the properties of MHD turbulence. To calculate synchrotron radiation intensity, we assume that these relativistic electrons obey a homogeneous and isotropic power-law distribution, as described below:
\begin{equation}
n(E){\rm d}E=CE^{2\alpha-1}{\rm d}E,
\end{equation}
where $n(E){\rm d}E$ denotes the number density of relativistic electrons in the energy range between $E$ and $E+{\rm d}E$, $C$ is a normalization constant, and $\alpha=(1-\gamma)/2$ is photon spectral index associated with electron spectral index $\gamma$. 

The synchrotron radiation intensity is given by \citep{Ginzburg1965}
\begin{equation}
I({\bm X}) \propto \int_{0}^{L}({B_{\rm x}^2({\bm X}, z)+B_{\rm y
}^2}({\bm X}, z))^{\frac{1-\alpha}{2}}{\rm d}z,
\end{equation}
where $B_{\rm x}$ and $B_{\rm y}$ are two components of magnetic field perpendicular to the line of sight (LOS), ${\bm X}=(x,y)$ denotes a two-dimensional vector in the plane of sky (POS), and $L$ is the length of emission region along the LOS.
When synchrotron radiation is linearly polarized, its intensity is calculated as:
\begin{equation}
P_{\rm int}(\bm X)=p_{0}I(\bm X),
\end{equation}
where $p_{0}=(3-3\alpha)/(5-3\alpha)$ is polarization degree. Based on this, Stokes parameters can be expressed as: $Q(\bm X)=P_{\rm int}(\bm X)\cos2\varphi_{0}$ and $U(\bm X)=P_{\rm int}(\bm X)\sin2\varphi_{0}$, with the polarization angle $\varphi_{0}=\pi/2+\arctan(B_{\rm y}/B_{\rm x})$. When Faraday rotation effect is taken into account, the above angle becomes
\begin{equation}
\varphi=\varphi_{0}+\lambda^{2}\phi, 
\end{equation}
where $\phi=0.81\int_{0}^{z}n_{\rm e}({\bm X},z^{'})B_{\parallel}(\bm X,z^{'})dz^{'}$ is Faraday rotation measure, $n_{\rm e}$ is the number density of thermal electron and $B_{\parallel}$ is the component of magnetic field along the LOS. Consequently, the Stokes parameters are modified to $Q(\bm X)=P_{\rm int}(\bm X)\cos2\varphi$ and $U(\bm X)=P_{\rm int}(\bm X)\sin2\varphi$. Based on these parameters, the SPI is calculated as
\begin{equation}
PI=\sqrt{Q^2+U^2}. \label{equ7}
\end{equation}

In addition to above two statistics, other synchrotron polarization diagnostics have been proposed to reveal the properties of MHD turbulence \citep{Gaensler2011, Herron2018, Zhang2019}. 
In this work, we adopt SPG, defined as:
\begin{equation}
P_{\rm grad}=\sqrt{(\frac{\partial Q}{\partial x})^2+(\frac{\partial U}{\partial x})^2+(\frac{\partial Q}{\partial y})^2+(\frac{\partial U}{\partial y})^2}. \label{equ8}
\end{equation}

\begin{deluxetable*}{cccccc}
\tabletypesize{\scriptsize}
\tablecaption{Different data cubes of compressible MHD turbulence.}
\tablewidth{170mm}
\tablehead{\colhead{Run} & \colhead{$B_{0}$} & \colhead{$M_{\rm A}$} 
   & \colhead{$M_{\rm s}$}
   & \colhead{$\beta$}
   & \colhead{Descriptions}}

\startdata
  1 & 2.0  & 0.34  & $\sim7.47$  & 0.004  & sub-Alfv\'enic and supersonic turbulence\\
  2 & 1.0  & 0.57  & $\sim7.47$  & 0.010  & sub-Alfv\'enic and supersonic turbulence\\
  3 & 0.1  & 1.53  & $\sim7.47$  & 0.108  & super-Alfv\'enic and supersonic turbulence\\
  4 & 0.05  & 2.55  & $\sim7.47$  & 0.242   & super-Alfv\'enic and supersonic turbulence\\ 
  5 & 1.0   & $\sim 0.58$  & 0.87  & 1.295  & sub-Alfv\'enic and subsonic turbulence\\
  6 &1.0  & $\sim 0.58$  & 3.16  & 0.067  & sub-Alfv\'enic and supersonic turbulence\\
  7 & 1.0  & $\sim 0.58$  & 6.78  & 0.012  & sub-Alfv\'enic and supersonic turbulence\\
  8 &1.0  & $\sim 0.58$  & 9.92  & 0.005  & sub-Alfv\'enic and supersonic turbulence\\
\enddata
\label{table:1}
\end{deluxetable*}

\subsection{Methods of Intermittency}
Intermittency, related to inhomogeneity of physical quantities, can be characterized by probability distribution function (PDF). 
However, this method only describes the distribution of 
fluctuations in physical quantity $Y(\bm X)$ at a specific scale, and this fluctuation is defined as:
\begin{equation}
\delta Y({\bm R})=Y({\bm {X}+\bm {R}})-Y({\bm X}).
\end{equation}
When this fluctuation is homogeneous, its distribution follows a Gaussian form.
By contrast, in the presence of inhomogeneity, the distribution exhibits non-Gaussian features with pronounced heavy tails.
These can be regarded as a criterion to identify intermittency.

In addition, the degree of intermittency can be quantified by evaluating the scaling exponent of multi-order structure function. 
In general, the multi-order structure function can be written as:
\begin{equation}
SF^{(p)}(\bm R)=\langle|\delta Y(\bm R)|^{p}\rangle,
\end{equation}
where $\langle...\rangle$ denotes the spatial average.
This function typically exhibits a certain dependence on separation scale $R$ within the inertial range, namely $SF^{(p)}(\bm R)\propto R^{\zeta(p)}$, where the scaling index $\zeta(p)$ is order-dependent.
A linear relationship between the scaling exponent $\zeta(p)$ and the order $p$ indicates the absence of intermittency, whereas the deviation from Kolmogorov scaling signifies the appearance of intermittency. 
In practice, most studies use the extended self-similarity (ESS) method \citep{Benzi1993}. It describes the scaling relation between multi-order and third-order structure functions and extends the analysis from the inertial range to the dissipative one. In this work, we employ the ESS method to obtain the relative scaling exponent ($\xi (p)$) between multi-order and third-order structure functions.

To characterize the deviation of scaling exponents from the theoretical model, the root-mean-square deviation (RMSD) is presented in this paper as a quantitative indicator, defined as follows.
\begin{equation}
\rm RMSD=\sqrt{\frac{1}{N}\sum_{i=1}^N(\xi_{i}^{\rm th}-\xi_{i}^{\rm mea})^2},
\end{equation}
where $\xi_{i}^{\rm th}$ denotes theoretical scaling exponent, $\xi_{i}^{\rm mea}$ represents fitted measurement value, and $N$ is the total number of samples.
A larger value indicates a more significant deviation from theoretical model. 

\section{Numerical simulation of MHD turbulence}\label{sec:simulation}
In this paper, we conduct numerical simulations by solving ideal single-fluid MHD equations with the third-order-accurate hybrid essentially non-oscillatory code \citep{cho2003MNRAS}.
The equations of MHD turbulence are written as:
\begin{gather}
\frac{\partial \rho }{\partial t} + \nabla \cdot (\rho {\bm v})=0, \label{eq:den}\\
\rho[\frac{\partial {\bm v}}{\partial t} + ({\bm v}\cdot \nabla) {\bm v}] +  \nabla p_{\rm th}- \frac{{\bm J} \times {\bm B}}{4\pi} ={\bm F}, \label{eq:vel}\\
\frac{{\partial {\bm B}}}{{\partial t}} -\nabla \times ({\bm v} \times{\bm B})=0,\label{eq:mag}\\
\nabla \cdot {\bm B}=0,
\end{gather}
where $t$ represents evolution time of MHD turbulence, $p_{\rm th}$ is thermal gas pressure, ${\bm J}=\nabla \times {\bm B}$ denotes current density, and $\bm F$ is random driving force. Note that these physical quantities are dimensionless. Periodic boundary conditions are applied to the data cube with a side length of 2$\pi$, and numerical simulation has a resolution of $512^3$.

The MHD turbulence is driven in a solenoidal driving force at the wavenumber of $k\approx 2.5$, with energy continuously injected into the system. For the numerical simulation, we set the initial magnetic field along the $x$ direction. The properties of MHD turbulence are characterized by two parameters: Alfv\'enic Mach number $M_{\rm A}=V_{\rm L}/V_{\rm A}$ and sonic Mach number $M_{\rm s}=V_{\rm L}/c_{\rm s}$. The former characterizes the level of magnetization, while the latter is used to quantify compressibility. The detailed information has been listed in Table \ref{table:1}.

\section{Numerical results}\label{sec:result}

To construct synthetic observations, we calculate the SPI and SPG via Eqs. (\ref{equ7}) and (\ref{equ8}) using the data cubes in Table \ref{table:1}. 
These dimensionless data cubes are parameterized using typical values of Galactic ISM. We adopt three key parameters: thermal electron number density of $n_{\rm e}=0.01~\rm cm^{-3}$, magnetic field of $B=1.23 ~\rm \mu G$, and spatial scale of $L=1000~\rm pc$.
When calculating the Faraday rotation measure, we assume thermal electron density $n_{\rm e}$ proportional to plasma density $\rho$.

\begin{figure*}[ht]
\center
\includegraphics[width=0.8\textwidth]{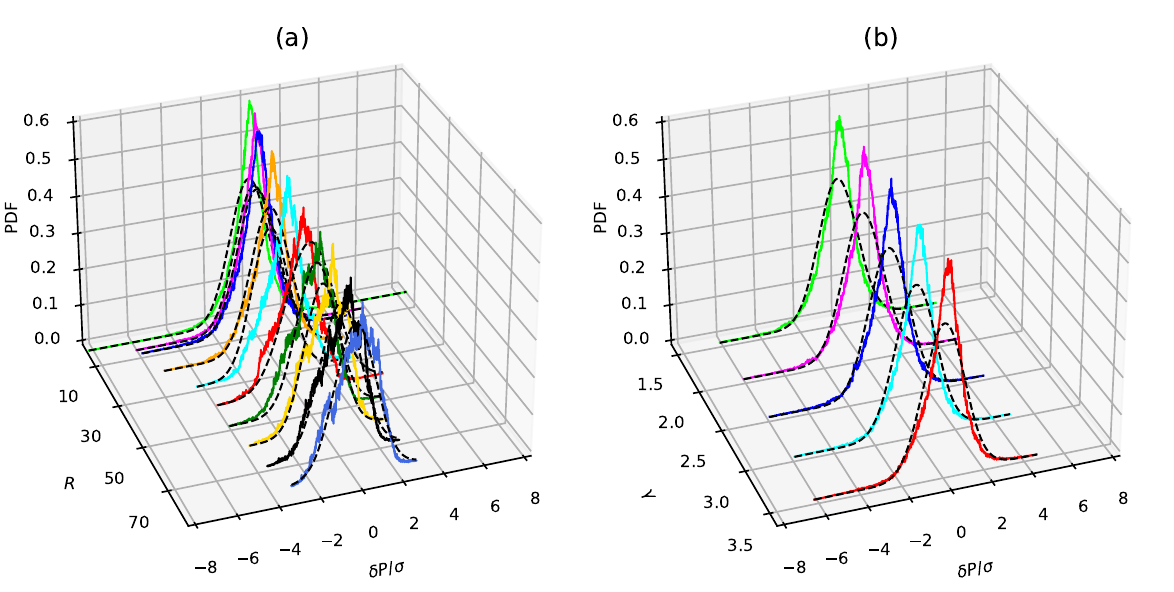}
\caption{PDFs of SPI fluctuations normalized by its standard deviation $\sigma$ at different separation scales $R$ (panel (a)) and electron spectral indices $\gamma$ (panel (b)) for the simulation of Run2. The black dashed lines represent Gaussian distributions.
}\label{fig:pdf} 
\end{figure*}

\begin{figure*}[ht]
\centering
\includegraphics[width=1.0\textwidth]{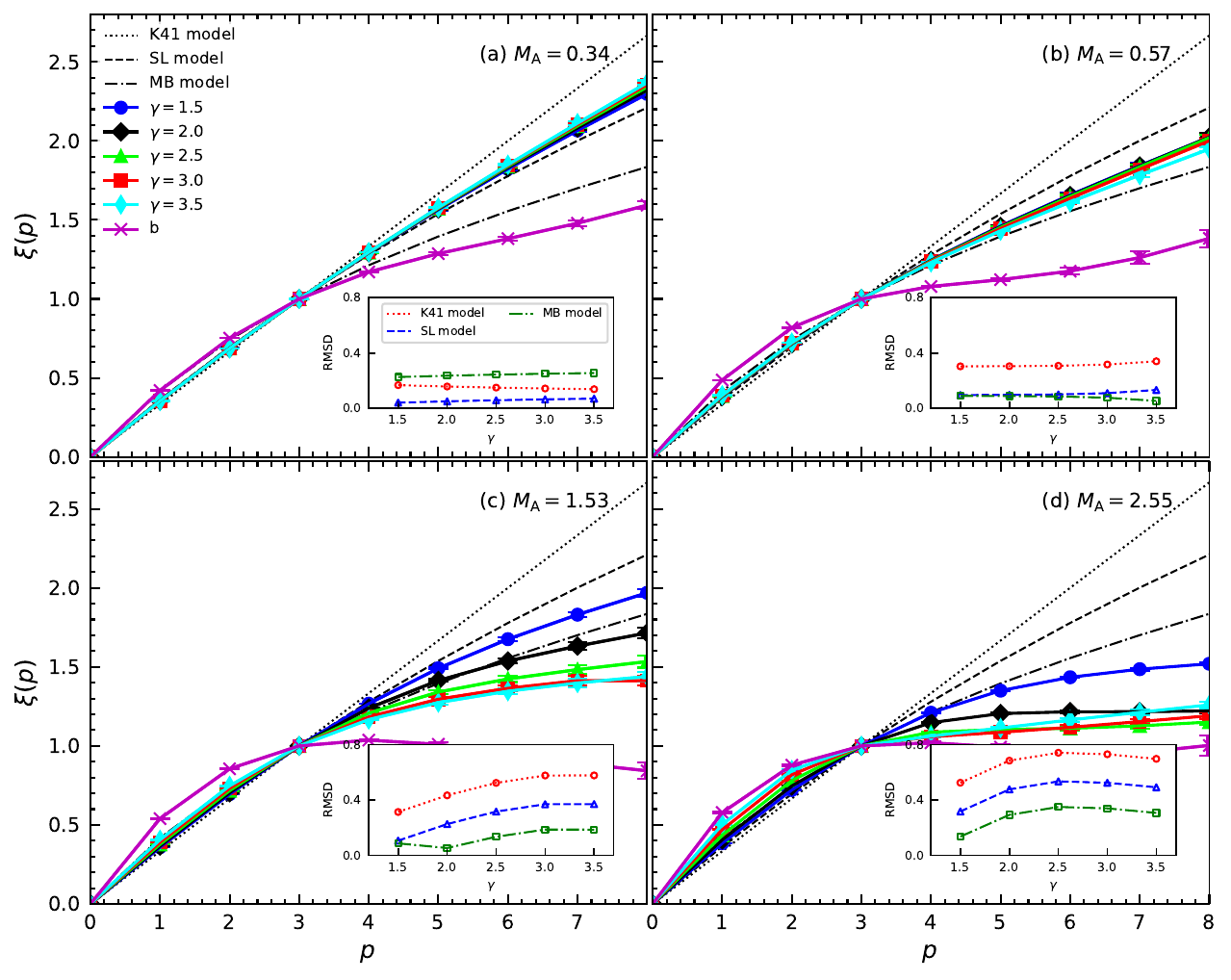}
\caption{Relative scaling exponents for SPI as a function of the order at different electron spectral indices $\gamma$ for four Alfv\'enic turbulence regimes with fixed $M_{\rm s}\sim 7.47$. The magenta lines with cross markers denote the distributions of relative scaling exponents for the 3D magnetic field.
The insets in panels (a)-(d) show RMSD value versus $\gamma$ corresponding to 
three theoretical models.}  
\label{fig:PI_different_alf} 
\end{figure*}

\subsection{Intermittency of Magnetized ISM Characterized by SPI}
\subsubsection{Effect of Electron Spectral Index}

Here, we investigate the influence of separation scale and electron spectral index on the intermittency of magnetized ISM by SPI.
The SPI is calculated using Run2 listed in Table \ref{table:1}, with the frequency of $\nu=10~ \rm GHz$. 
To qualitatively describe intermittency, Fig. \ref{fig:pdf} presents the PDFs of SPI fluctuations normalized by their standard deviation at different separation scales and electron spectral indices, corresponding to panels (a) and (b), respectively. 
It can be seen from Fig. \ref{fig:pdf}(a) that the PDFs of SPI fluctuations deviate from a Gaussian distribution over the range of separation scales, and the deviation exhibits a clear scale dependence: the deviation becomes more significant at smaller scales, while it weakens at larger scales. This means that 
the intermittency of magnetized ISM associates with the scale, and this becomes abundant at small scales.
In Fig. \ref{fig:pdf}(b), all PDFs of SPI fluctuations deviate from Gaussian profile, and the degree of deviation is nearly identical at different electron spectral indices. 
This indicates that the intermittency of magnetized ISM does not depend on the electron spectral index in the sub-Alfv\'enic and supersonic turbulence regime.

\begin{figure*}[ht]
\centering
\includegraphics[width=1.0\textwidth]{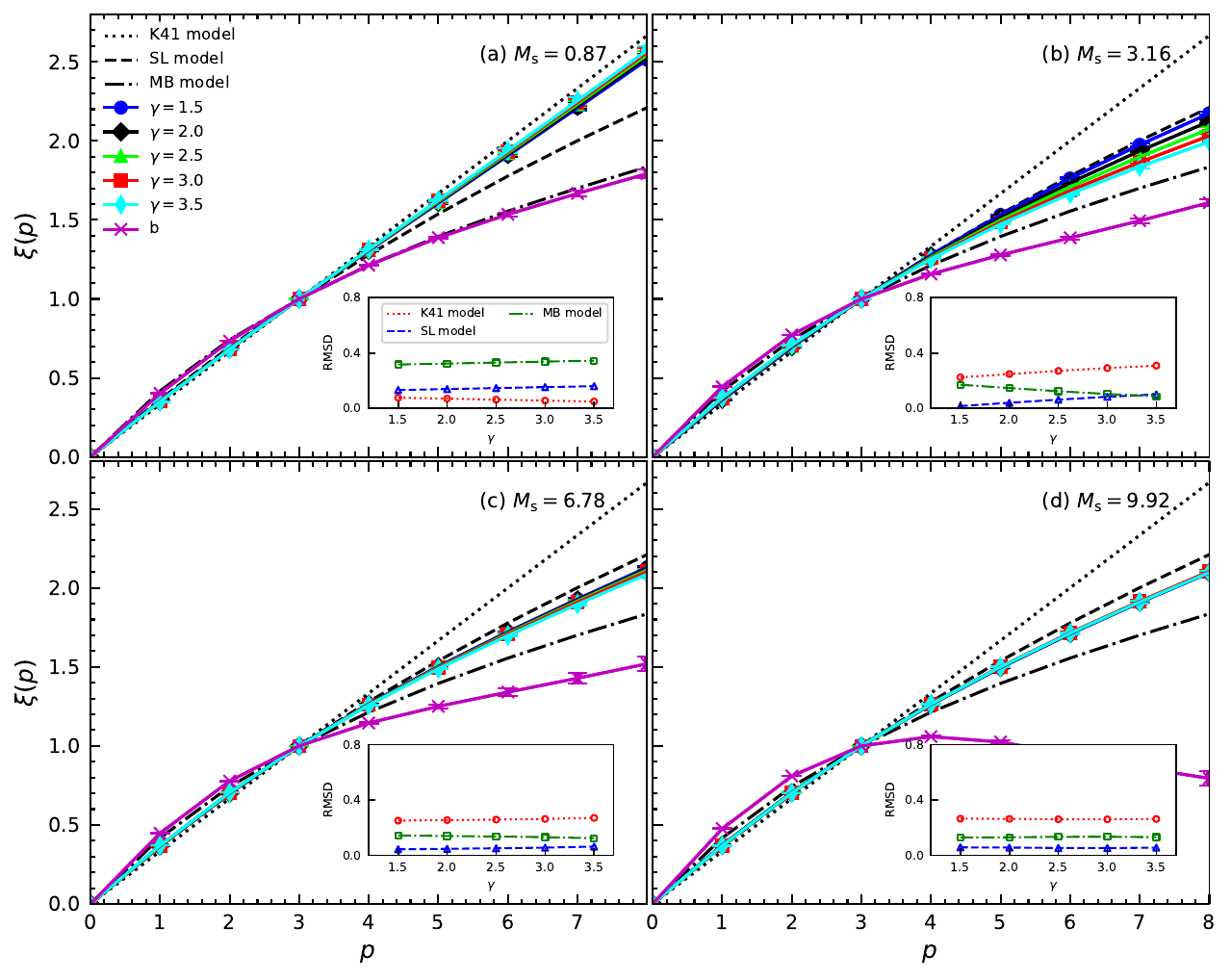}
\caption{Relative scaling exponents for SPI as a function of the order at different electron spectral indices $\gamma$ for four sonic turbulence regimes with fixed $M_{\rm A}\sim 0.58$. The magenta lines with cross markers denote the distributions of relative scaling exponents for the 3D magnetic field.
The insets in panels (a)-(d) show RMSD value versus $\gamma$ corresponding to 
three theoretical models.}   
\label{fig:PI_different_sonic} 
\end{figure*}

To further explore the dependence on electron spectral index, we analyze this behavior by the relative scaling exponent in different Alfv\'enic turbulence regimes, as presented in Fig. \ref{fig:PI_different_alf}.
The error bar represents the standard derivation.
From this figure, it is evident that the profiles of relative scaling exponent for SPI become almost unchanged at different electron spectral indices for the sub-Alfv\'enic turbulence. 
In contrast, the distributions of relative scaling exponent behave differently from $\gamma=1.5$ to $3.5$ for the super-Alfv\'enic turbulence. As the electron spectral index increases, the change of $\xi(p)$ with $p$ gradually deviates from MB model, indicating enhanced intermittency. 
This finding is also clearly presented in the inset. In the sub-Alfv\'enic turbulence regime, the RMSD value remains unchanged as the electron spectral index increases, while this value increases in the super-Alfv\'enic turbulence regime. 
This indicates that the intermittency of magnetized ISM increases with the electron spectral index only for super-Alfv\'enic turbulence.

We also visualize how the intermittency of magnetized ISM changes with Alfv\'enic Mach numbers in Fig. \ref{fig:PI_different_alf}. Comparing all four scenarios, the curves of relative scaling exponent versus order depart markedly from K41 model with increasing Alfv\'enic Mach numbers, suggesting more pronounced intermittency. 
This is because super-Alfv\'enic turbulence with a weak magnetic field facilitates the generation of strong fluctuations.
The result above is also validated in the inset. The RMSD values corresponding to K41 model tend to increase 
in the super-Alfv\'enic turbulence regime, demonstrating enhanced intermittency.
This also suggests that magnetic field plays an important role in forming high-dimensional structures in supersonic turbulence.

\begin{figure}[htbp]
\centering
\includegraphics[width=0.45\textwidth]{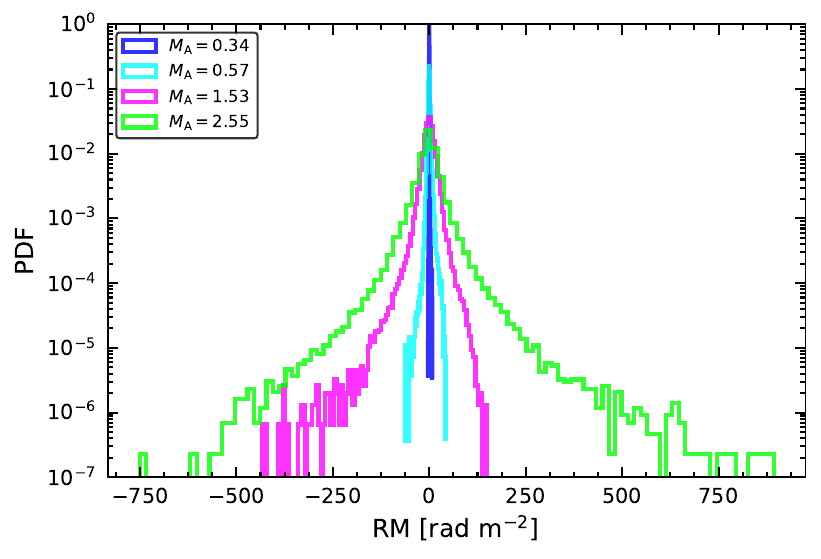}
\caption{The PDFs of Faraday rotation measure at different Alfv\'enic Mach numbers for fixed $M_{\rm s}\sim 7.47$.
}\label{fig:map_faraday_alf} 
\end{figure}

\begin{figure}[htbp]
\centering
\includegraphics[width=0.45\textwidth]{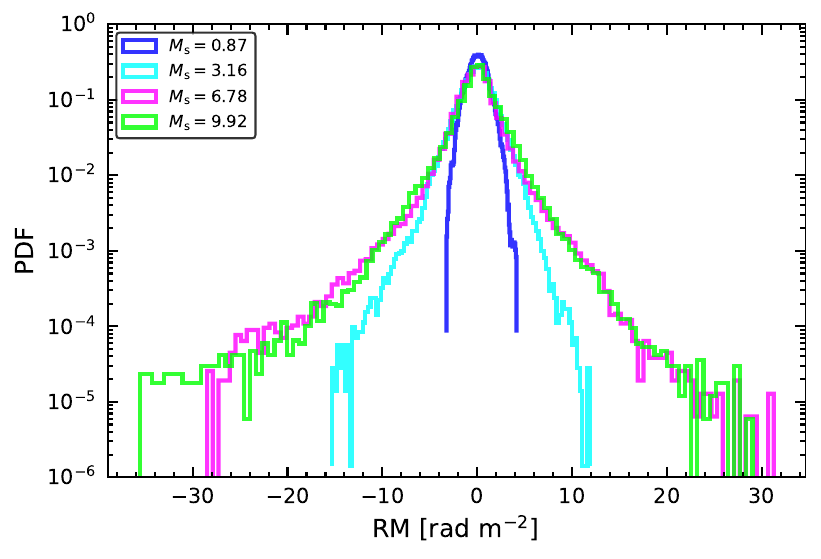}
\caption{The PDFs of Faraday rotation measure at different sonic Mach numbers for fixed $M_{\rm A}\sim 0.58$.
}\label{fig:map_faraday_sonic} 
\end{figure}

\begin{figure*}[ht]
\centering
\includegraphics[width=1.0\textwidth]{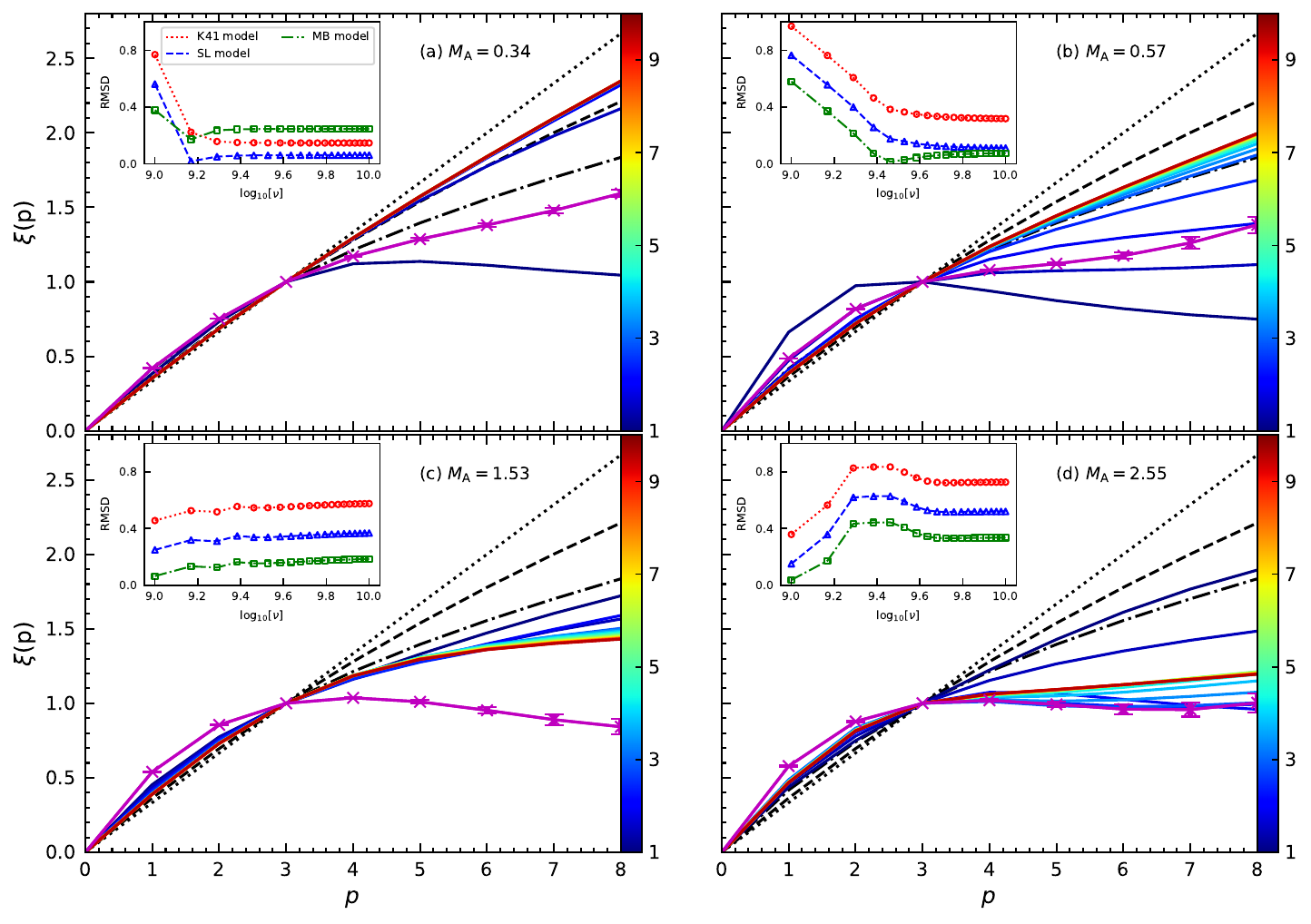}
\caption{Relation between relative scaling exponent and order for SPI calculated at individual frequencies for four Alfv\'enic turbulence regimes with fixed $M_{\rm s}\sim 7.47$. The dotted, dashed, and dashed-dotted lines correspond to K41, SL, and MB models, respectively. The magenta lines with cross markers denote the distributions of relative scaling exponents for the 3D magnetic field.
The color bar indicates the change in frequency in units of 1 GHz. 
The insets in panels (a)-(d) show RMSD values versus frequency corresponding to three theoretical models.
}\label{fig:SPI_frequency_alf} 
\end{figure*}

\begin{figure*}[ht]
\centering
\includegraphics[width=1.0\textwidth]{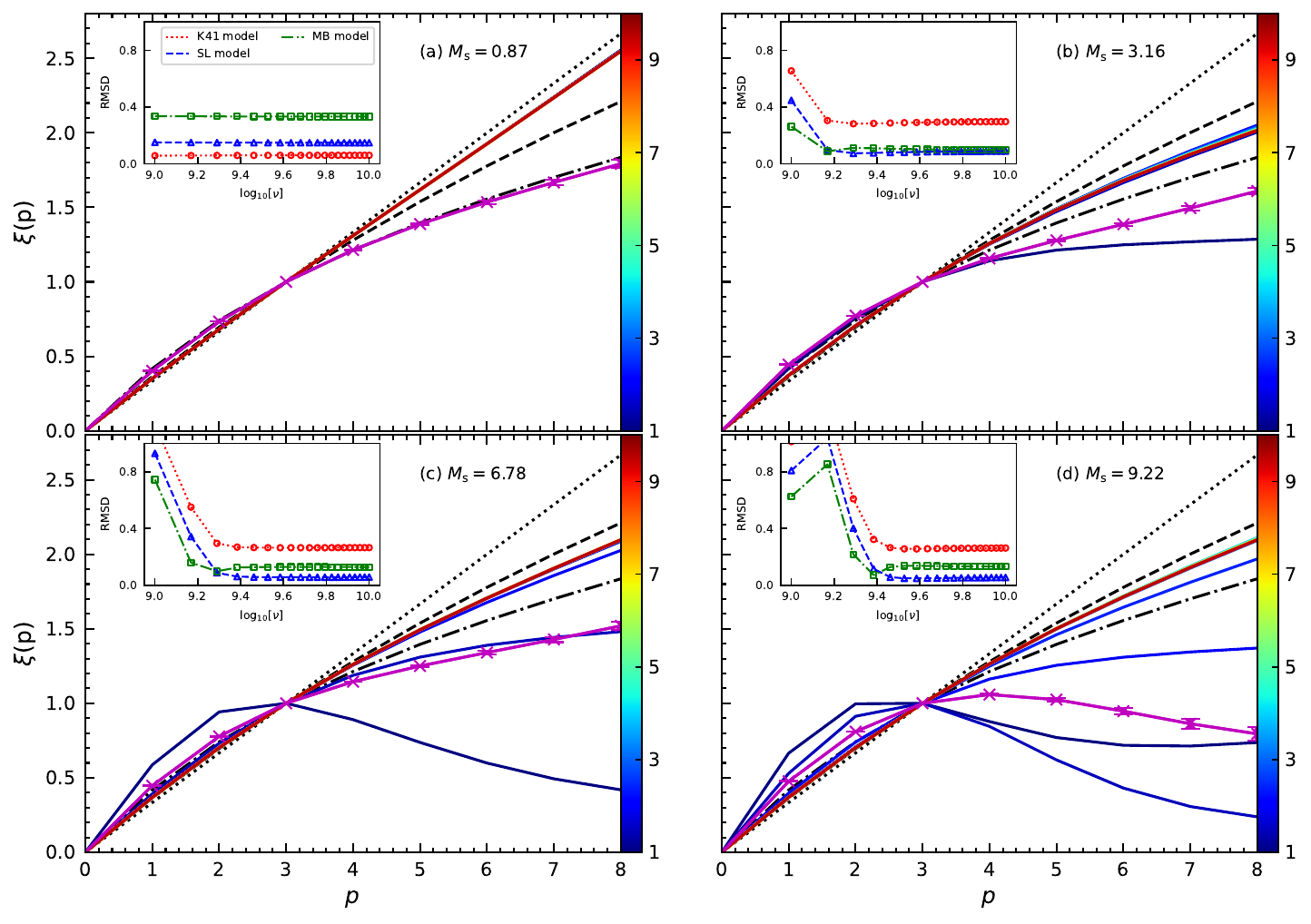}
\caption{Relation between relative scaling exponent and order for SPI calculated at individual frequencies for four sonic turbulence regimes with fixed $M_{\rm A}\sim 0.58$. The dotted, dashed, and dashed-dotted lines correspond to K41, SL, and MB models, respectively. The magenta lines with cross markers denote the distributions of relative scaling exponents for the 3D magnetic field.
The color bar indicates the change in frequency in units of 1 GHz. 
The insets in panels (a)-(d) show RMSD values versus frequency corresponding to three theoretical models.
}\label{fig:SPI_frequency_sonic} 
\end{figure*}

Similar to Fig. \ref{fig:PI_different_alf}, Fig. \ref{fig:PI_different_sonic} explores the dependence of intermittency on electron spectral index in different sonic turbulence regimes. 
In each panel, the profiles of relative scaling exponent behave similarly at different electron spectral indices. The RMSD values in the inset remain nearly unchanged as the electron spectral index increases. 
This finding suggests that there is no visible dependence on $\gamma$, regardless of whether MHD turbulence is subsonic or supersonic. 
It is not difficult to understand this phenomenon, because all relevant processes take place in the sub-Alfv\'enic turbulence regime, thus showing no dependence on $\gamma$.
This further confirms that the intermittency is nearly independent on $\gamma$ for strongly magnetized turbulence.

Furthermore, Fig. \ref{fig:PI_different_sonic} illustrates the changes of relative scaling exponent with sonic Mach number.  
From panels (a)-(d), it can be seen that the profiles of relative scaling exponent approach K41 model in the subsonic turbulence regimes, while those are almost consistent with SL model in the supersonic one. 
It is evident in the inset that the RMSD values converge toward zero for subsonic and supersonic turbulence, but these two cases correspond to K41 and SL models, respectively.
This reveals that sonic Mach number affects the intermittency of magnetized ISM: the larger the sonic Mach number, the more pronounced the intermittency. The reason is the formation of shocks in supersonic turbulence to enhance intermittency.

In the two figures above, we also present the relative scaling exponents of the 3D magnetic field, as indicated by the magenta lines with cross markers. These figures show that the relative scaling exponents for the 3D magnetic field deviate more substantially from the K41 model than those for SPI obtained at different electron spectral indices, regardless of the Alfv\'enic and sonic Mach numbers. This result indicates that the intermittency derived from SPI is weaker than the intrinsic intermittency of the underlying 3D magnetic field. This suppression primarily arises from line-of-sight integration: synchrotron polarized emission accumulates contributions along the entire line of sight, where sharp intermittent spatial structures are smoothed by spatial averaging. Notably, the above analysis applies to SPI at 10 GHz, where Faraday depolarization effect remains weak.

\subsubsection{Strong and Weak Faraday Depolarization}

Fig. \ref{fig:map_faraday_alf} shows the PDFs of Faraday rotation measure at different Alfv\'enic Mach numbers. Evidently, all profiles are non-Gaussian.
The Faraday rotation measure ranges differently for each Alfv\'enic Mach number: from -3 to 8 for $M_{\rm A}=0.34$, -63 to 42 for $M_{\rm A}=0.57$, -432 to 149 for $M_{\rm A}=1.53$, and -750 to 893 for $M_{\rm A}=2.55$.
This indicates that Faraday rotation measure gradually increases as the Alfv\'enic Mach number increases.
This phenomenon may result from the weak magnetic field, which further reduces the suppression on thermal electron density fluctuations.
Similar to Fig. \ref{fig:map_faraday_alf}, Fig. \ref{fig:map_faraday_sonic} exhibits the PDFs of Faraday rotation measure at different sonic Mach numbers.
As the sonic Mach number increases, the PDFs gradually develop more distinct heavy-tailed features.
The corresponding values cover different ranges: from -3 to 4 for $M_{\rm s}=0.87$, -15 to 12 for $M_{\rm s}=3.16$, -28 to 31 for $M_{\rm s}=6.78$ and -35 to 29 for $M_{\rm s}=9.92$. 
For supersonic turbulence, this behavior is mainly attributed to shocks.

Fig. \ref{fig:SPI_frequency_alf} presents the relative scaling exponent for SPI in the frequency range of $1$ to $10~\rm GHz$ at different Alfv\'enic Mach numbers. 
In panels (a) and (b), the deviation of relative scaling exponent from K41 model is more pronounced at low frequencies than at high frequencies. This may be attributed to the dominant role of strong depolarization at low frequencies, generating more coherent structures.
In panels (c) and (d), there is the opposite behavior.
For super-Alfv\'enic turbulence, Faraday depolarization and intrinsic synchrotron emission jointly suppress the formation of inhomogeneous structures at low frequencies.
The embedded subplots further confirm the above conclusion. The RMSD values decrease with frequency in sub-Alfv\'enic turbulence, while the opposite trend appears in super-Alfv\'enic turbulence. In both cases, the RMSD values converge to a constant at high frequencies due to weak Faraday depolarization.

\begin{figure*}[ht]
\centering
\includegraphics[width=0.8\textwidth]{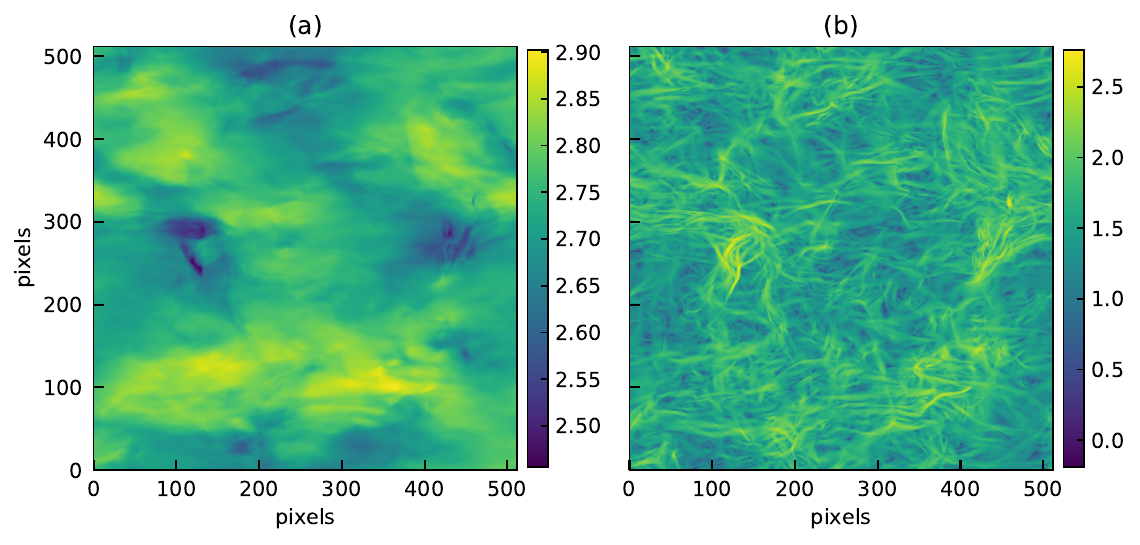}
\caption{2D structure maps of SPI (panel (a)) and SPG (panel (b)) based on Run2 listed in Table \ref{table:1}. 
}\label{fig:SPI_SPG_compare} 
\end{figure*}

In the same frequency range from $1$ to $10~\rm GHz$, Fig. \ref{fig:SPI_frequency_sonic} explores the relative scaling exponents for SPI at different sonic Mach numbers. In panel (a), the profiles of relative scaling exponents coincide at all frequencies and nearly agree with the K41 model. Panels (b)-(d) illustrate that the changes of relative scaling exponent with order gradually converge toward the K41 model with rising frequency, and their values lie between the SL model and MB model.
This can be reflected in the inset. Except for panel (a), 
the RMSD values corresponding to SL model first decrease and then saturate at 0.1 with increasing frequency.
This phenomenon is explained as follows.
For panel (a), subsonic turbulence prevents shock formation, and the magnetic field therefore efficiently constrains plasma motion, producing less coherent structures.
For other cases, strong Faraday depolarization at low frequencies amplifies intermittency.
At high frequencies, the Faraday depolarization weakens substantially, with intrinsic synchrotron emission governing the intermittency level and yielding reduced intermittency.
Furthermore, the frequency dependence of relative scaling exponent becomes more pronounced at higher sonic Mach numbers.
These results are also supported by the embedded subplots. 
As the sonic Mach number increases, the turning frequency corresponding to the change from large to small RMSD values shifts towards higher frequencies.
This further confirms that the intermittency of magnetized ISM exhibits stronger frequency dependence in highly compressible turbulence.

We further compare the intermittency of the 3D magnetic field and SPI, with the relative scaling exponents of the former shown by the magenta lines with cross markers in Figs. \ref{fig:SPI_frequency_alf} and \ref{fig:SPI_frequency_sonic}.
These figures demonstrate that the intermittency of the 3D magnetic field is stronger than that of SPI at high frequencies, while the opposite holds at low frequencies. This reversal occurs only for the sub-Alfv\'enic and supersonic turbulence regime. This may be because supersonic turbulence generates strong density fluctuations, enhancing the Faraday depolarization effect at low frequencies so that it exceeds the contribution from intrinsic synchrotron emission. Nevertheless, for both the sub-Alfv\'enic/subsonic and super-Alfv\'enic/supersonic turbulence regimes, the intermittency of the 3D magnetic field is stronger than that of SPI for all frequencies considered in this work.

\begin{figure*}[ht]
\centering
\includegraphics[width=0.8\textwidth]{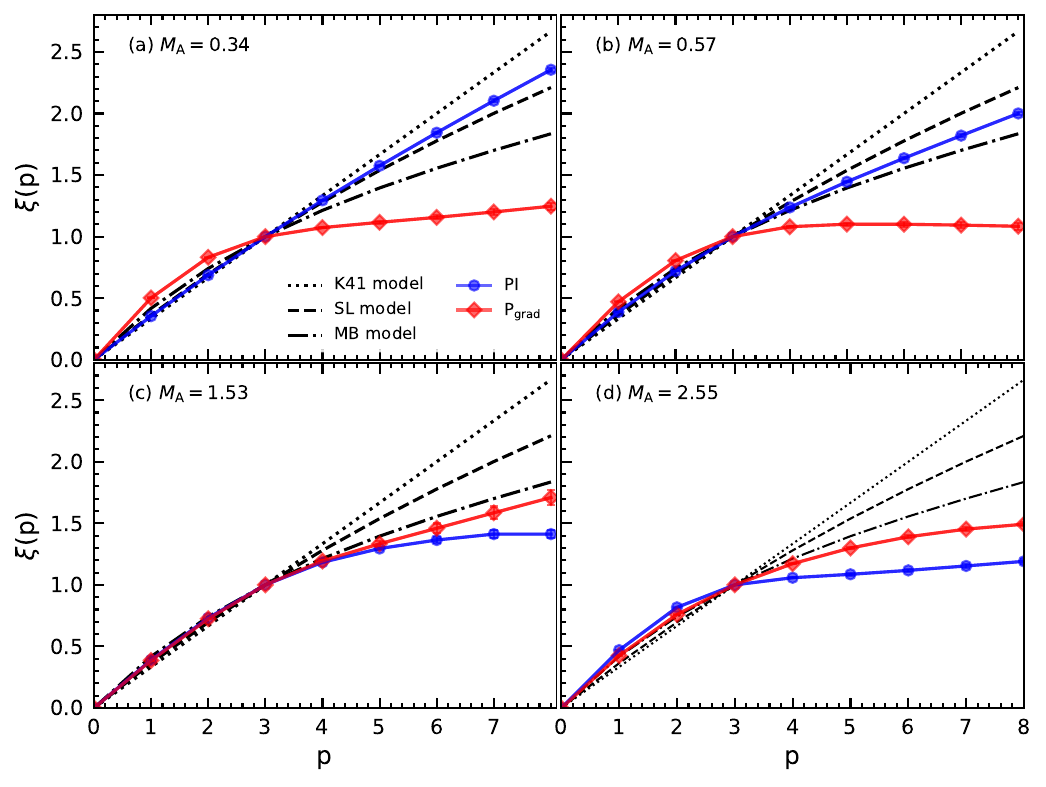}
\caption{Relative scaling exponents for SPI (blue lines) and SPG (red lines) as a function of the order at different Alfv\'enic Mach numbers with fixed $M_{\rm s}\sim 7.47$, shown in panels (a)-(d), respectively. 
}\label{fig:SPI_SPG_compare_scaling_alf}
\end{figure*}

\begin{figure*}[ht]
\centering
\includegraphics[width=0.8\textwidth]{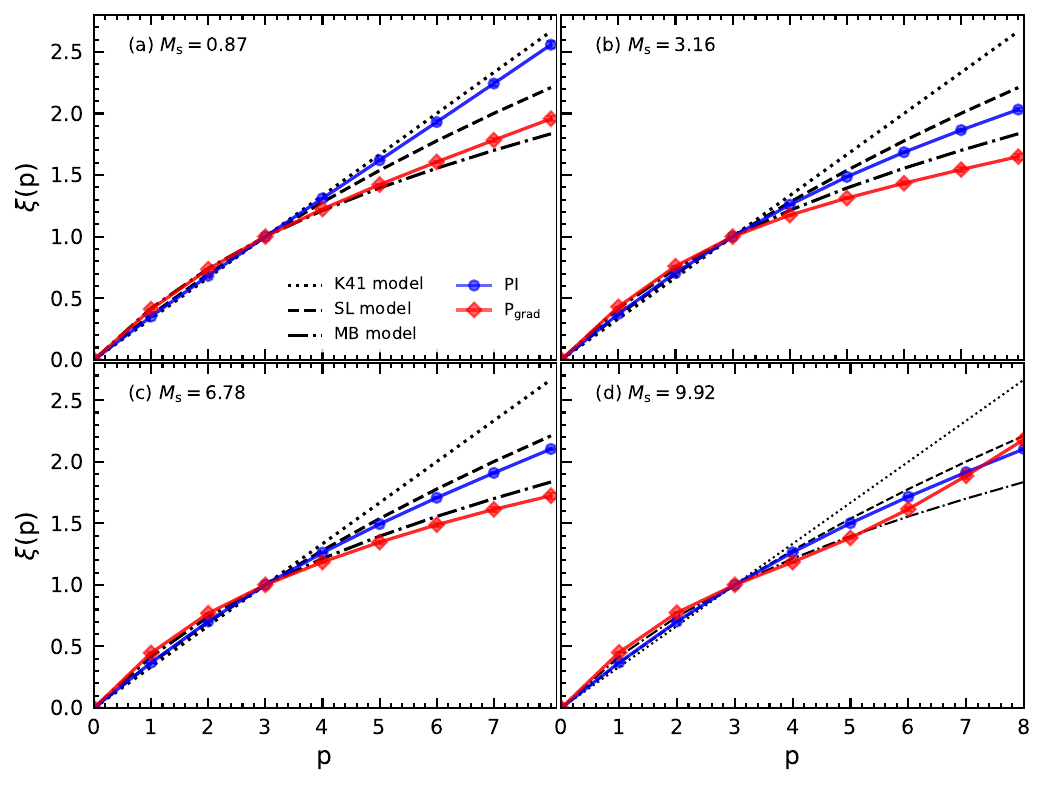}
\caption{Relative scaling exponents for SPI (blue lines) and SPG (red lines) as a function of the order at different sonic Mach numbers with fixed $M_{\rm A}\sim0.58$, shown in panels (a)-(d), respectively.
}\label{fig:SPI_SPG_compare_scaling_sonic} 
\end{figure*}

\subsection{Intermittency of Magnetized ISM Characterized by SPG} 

\subsubsection{Intermittency of Magnetized ISM: SPI vs. SPG}
Based on Run 2 listed in Table \ref{table:1}, we calculate the SPI and SPG with the electron spectral index of $\gamma=3$ and frequency of $\nu=10~\rm GHz$. We obtain 2D structure maps of SPI and SPG, shown in panels (a) and (b) of Fig. \ref{fig:SPI_SPG_compare}, respectively. 
It can be clearly seen from these two panels that the structures of SPI and SPG extend along the $x$ axis, indicating the orientation of mean magnetic field.
Compared with SPI, the image of SPG exhibits finer filamentary structures with higher peaks. 
This qualitatively demonstrates that SPG has more pronounced intermittency than SPI.
From the color bar, we can clearly see that the overall value range of SPI is narrower than that of SPG, which stems from amplified gradient fluctuations.

To further compare the intermittency of magnetized ISM characterized by SPI and SPG, we present the corresponding relative scaling exponents at different Alfv\'enic Mach numbers in Fig. \ref{fig:SPI_SPG_compare_scaling_alf}. In the sub-Alfv\'enic turbulence regime, the relative scaling exponents for SPG deviate more significantly from K41 model than those for SPI. In contrast, the opposite behavior is displayed in the super-Alfv\'enic turbulence regime. 
This suggests that the relative intermittency level characterized by SPI and SPG reverses in different Alfv\'enic turbulence regimes. This phenomenon can be explained as follows.
In the sub-Alfv\'enic turbulence regime, SPG is capable of distinguishing more small-scale coherent structures compared with SPI. 
In the super-Alfv\'enic turbulence regime, the calculation of SPG introduces numerical noise that masks signatures of some small-scale coherent structures, leading to a reduction in intermittency.
Similarly, we investigate the intermittency differences between SPI and SPG for magnetized ISM at different sonic Mach numbers, as shown in Fig. \ref{fig:SPI_SPG_compare_scaling_sonic}.
It is evident that the relative scaling exponents for SPG show a larger deviation from K41 model compared with those for SPI in four panels. This indicates that the SPG-characterized intermittency of magnetized ISM is more abundant than the SPI-characterized one. Furthermore, as the sonic Mach number increases, the relative scaling exponents for SPG depart more obviously from K41 model, except for $M_{\rm s}=9.92$.

\begin{figure*}[ht]
\centering
\includegraphics[width=1.0\textwidth]{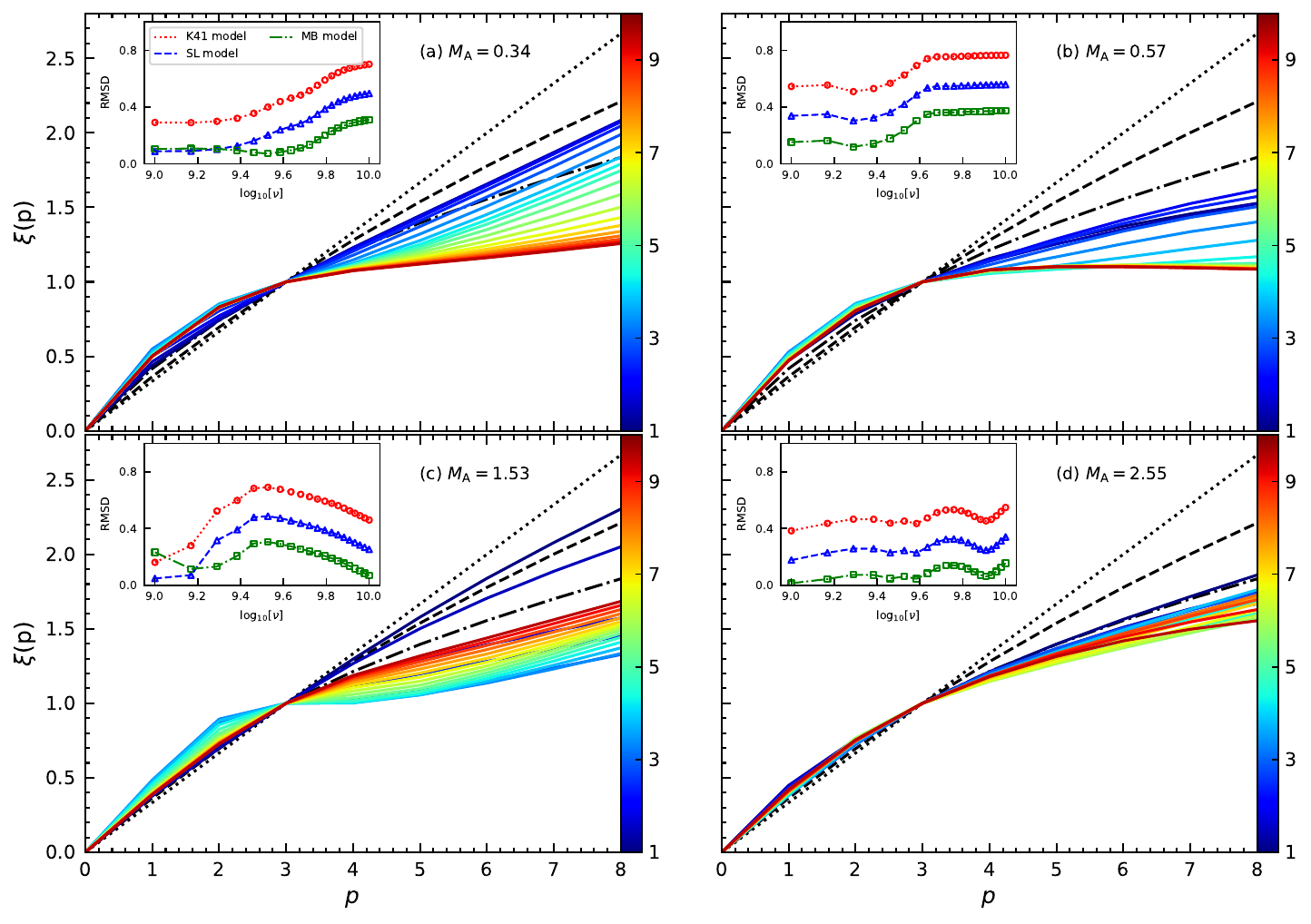}
\caption{Relative scaling exponent for SPG versus order calculated at individual frequencies for four Alfv\'enic turbulence regimes with fixed $M_{\rm s}\sim 7.47$.  The color bar indicates the change in frequency in units of 1 GHz. The insets illustrate the relationship between RMSD values and frequency, corresponding to K41, SL and MB models, respectively.
}\label{fig:SPG_scaling_alf} 
\end{figure*}

\begin{figure*}[ht]
\centering
\includegraphics[width=1.0\textwidth]{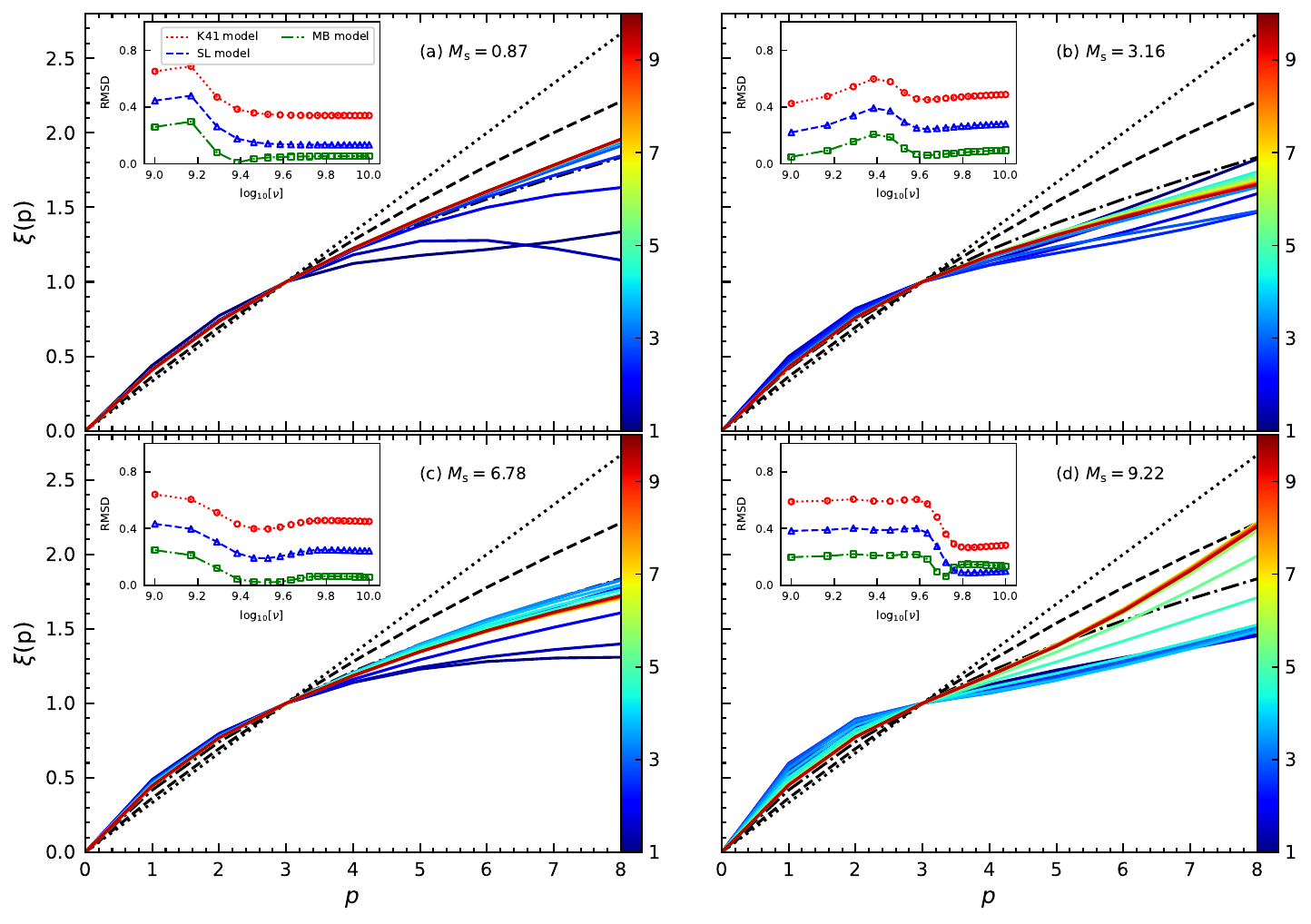}
\caption{Relative scaling exponent for SPG versus order calculated at individual frequencies for four sonic turbulence regimes with fixed $M_{\rm A}\sim 0.58$. The color bar indicates the change in frequency in units of 1 GHz. The insets illustrate the relationship between RMSD values and frequency, corresponding to K41, SL and MB models, respectively.
}\label{fig:SPG_scaling_sonic} 
\end{figure*}

\subsubsection{Strong and Weak Faraday Depolarization}

Fig. \ref{fig:SPG_scaling_alf} shows the relationship between the relative scaling exponent for SPG and the order for each frequency at four Alfv\'enic Mach numbers. 
For sub-Alfv\'enic turbulence, the slopes between relative scaling exponent and order become shallower than those of SL model as the frequency increases, revealing the generation of structures with dimensions higher than 1. 
For super-Alfv\'enic turbulence, the relative scaling exponents vary with frequency, and no distinct relationship exists between them.
This feature can also be seen in the embedded subplot. The RMSD values exhibit a gradual upward trend with increasing frequency in the sub-Alfv\'enic turbulence regime, while no clear relationship is observed in the super-Alfv\'enic turbulence regime. 
This behavior can be explained as follows. 
In sub-Alfv\'enic turbulence, the numerical noise generated by SPG removes partial small-scale coherent structures at low frequencies.
In super-Alfv\'enic turbulence, intrinsic synchrotron emission is essentially random. Combined with Faraday depolarization and noise, the structures become highly disordered with no obvious regularity.

Similarly, we investigate the changes of relative scaling index for SPG with the order at different sonic Mach numbers, with the results shown in Fig. \ref{fig:SPG_scaling_sonic}. 
From this figure, the relative scaling index becomes steeper and converges toward MB model with increasing frequency. The embedded subplot indicates that RMSD values are larger at low frequencies than at high frequencies. 
This may be due to the dominance of strong Faraday depolarization.
In addition, the distributions of RMSD value shift toward higher frequency as the sonic Mach number increases. 
This phenomenon is because the turbulence in larger sonic Mach numbers produces more shocks and influences Faraday rotation effect.

\section{Discussion}\label{sec:discussion}

Previous studies have investigated the effects of electron spectral index on power spectrum and anisotropy of MHD turbulence via SPI. This parameter merely changes the amplitude of spectrum, with little effect on its scaling index \citep{Lazarian2012,zhang2016, Lee2016} and anisotropy \citep{Herron2016, wang2020}.
This work complements existing studies on how the electron spectral index affects the properties of magnetized ISM characterized by SPI, by focusing on intermittency.
Our analysis suggests that the SPI-characterized intermittency of magnetized ISM is related to the electron spectral index under weak magnetization, while this relationship no longer holds for strong magnetization. 
This is one of the important conclusions of this paper. 
Accordingly, the electron spectral index should be taken into account when predicting the intermittency of magnetized ISM.

The dependence of MHD turbulence intermittency on Mach numbers has been studied by \cite{Wang2024}, whose results indicate that stronger intermittency appears when plasma parameters deviate substantially from unity. Different from the above work, we adopt a controlled-variable method. By fixing either sonic or Alfv\'enic Mach numbers, we separately explore the influence of the other Mach number on intermittency. 
The results indicate that both Alfv\'enic and sonic Mach numbers affect the intermittency of magnetized ISM, with this effect being more pronounced at higher Mach numbers.
Furthermore, the Mach number modulates how the intermittency depends on other physical parameters, highlighting its key role among all relevant influencing factors.

The SPI-characterized intermittency of magnetized ISM has a dependence on frequency, as verified in \cite{Wang2024}. However, this work adopted only four discrete frequency points for a single turbulence regime. In present work, we increase the number of sampled frequency points and investigate this dependence for four turbulence regimes. The results reveal that frequency dependence differs markedly between the sub-Alfv\'enic and super-Alfv\'enic turbulence regimes, shown in Figs. \ref{fig:SPI_frequency_alf} and \ref{fig:SPI_frequency_sonic}. 
Furthermore, the SPI cannot reliably characterize the intermittency of magnetized ISM at low frequencies. Unlike SPI, the intermittency of magnetized ISM characterized by SPG has frequency dependence, yet no obvious correlation can be observed.
It is worth noting that this method exhibits good robustness for intermittency measurements at low frequencies.
Currently, the available data are insufficient for a comprehensive investigation of this issue.
In fact, the driving mechanism and the angle between mean magnetic field and LOS are also important factors influencing the intermittency of magnetized ISM characterized by SPI or SPG, which will be further explored in future.

The SPG was originally proposed to estimate the sonic Mach number of interstellar gas, obtaining $M_{\rm s}\lessapprox 2$ for the warm ionized medium \citep{Gaensler2011}. 
Subsequently, this technique was extended to measure magnetic field \citep{Carmo2020, Lazarian2018, Zhang2020} and reveal the spectral properties \citep{Zhang2025}. 
This paper investigates the intermittency of magnetized ISM by SPG and validates its feasibility.
Compared with SPI, the SPG can capture finer structures, improving the robustness in characterizing small-scale coherent structures. 
This technique is applicable to measurement of intermittency in smaller scale and provides superior resolution of small-scale structures.
It is expected that the SPG can be applied to real radio observational data to explore the intermittency of magnetized ISM.

As a diagnostic for investigating magnetic field intermittency, SPI cannot fully recover the true intermittency level of the 3D magnetic field due to projection effects. At high frequencies, SPI is only sensitive to the intermittency of the plane-of-sky magnetic field $B_{\perp}$ (losing information on $B_{\parallel}$ fluctuations), while it can capture the intermittency of $B_{\parallel}$, $B_{\perp}$, $n_{\rm e}$ at low frequencies. Beyond these limitations, both Faraday-induced internal depolarization and beam smoothing from observations can further affect the measured intermittency. Given the inherent limitations of SPI, intermittency measured by SPI at high frequencies provides a lower limit for the true intermittency of 3D turbulent magnetic fields.

To characterize the intermittency level of magnetized ISM, we adopt PDF and scaling exponents of multi-order structure function as primary statistical diagnostics. The former can qualitatively analyze the intermittency at a fixed spatial scale, while the latter can offer
a quantitative description for the scaling exponent of high-order structure function. 
In general, the level of intermittency is evaluated by comparison with theoretical models based on the degree of deviation. Nevertheless, the deviations cannot be identified intuitively, as shown in Fig.  \ref{fig:SPG_scaling_alf}. For this reason, we introduce a method, namely the RMSD, to precisely quantify the departure of scaling exponents from their theoretical values. This method effectively characterizes the level of intermittency and has distinct advantages for investigating its dependence on the electron spectral index and frequency. At different electron spectral indices, the distributions of relative scaling exponents exhibit overlap, making it difficult to directly assess the degree of deviations. The RMSD values provide a quantitative description of these departures from theoretical models, enabling a quantitative judgment of the level of intermittency, as illustrated in Figs. \ref{fig:PI_different_alf} and \ref{fig:PI_different_sonic}. At different frequencies, the distributions of relative scaling exponent are also not readily discernible, yet the RMSD can clearly depict its frequency-dependent behavior, as shown in Figs. \ref{fig:SPI_frequency_alf}, \ref{fig:SPI_frequency_sonic}, \ref{fig:SPG_scaling_alf}, \ref{fig:SPG_scaling_sonic}. These results further validate the robustness of the RMSD as a metric for characterizing intermittency.
It should be noted that this method has some limitations. First, fitting range for the scaling exponents must be properly chosen. Second, low-resolution simulation data reduce the accuracy of fitted scaling exponents. 
In addition, current theoretical models remain inadequate to uncover the existing intermittency phenomena.

\section{Summary}\label{sec:summary}

In this paper, we qualitatively investigate the effects of electron spectral distribution, fluid compressibility, magnetization, and Faraday depolarization on the reconstruction of intermittency in magnetized ISM via SPI and SPG methods. The main findings are summarized as follows.
\begin{enumerate}[wide,labelwidth=!,labelindent=0.3pt]

\item Electron spectral distribution influences the measurement of intermittency in magnetized ISM, especially for environments dominated by random magnetic fields.

\item Fluid compressibility and magnetization enhance the ISM intermittency, which is particularly prominent in super-Alfv\'enic and supersonic turbulence regimes.

\item The SPG method offers a novel approach to probing the intermittency of magnetized ISM under strong Faraday depolarization, whereas the SPI method is only applicable for weak Faraday depolarization conditions. Moreover, SPG is capable of recovering the small-scale spatial characteristics of ISM intermittency.
\end{enumerate}

\begin{acknowledgments}

We would like to thank the anonymous referee for the valuable comments that improved our manuscript. R.Y.W. is grateful for the support from the National Natural Science Foundation of China (grant No. 12503028), the Guangxi Natural Science Foundation (No. 2026GXNSFBA00640012), Guangxi Young Elite Scientist Sponsorship Program (GXYESS2026007) and the Scientific Research Project of Guangxi Minzu University (No. 2024KJQD219).
J.F.Z. is grateful for the support from the National Natural Science Foundation of China (grant No. 12473046). 
\end{acknowledgments}

\bibliography{me}
\bibliographystyle{aasjournal}

\end{document}